\pdfoutput=1

\documentclass[11pt]{article}

\usepackage[final]{acl}
\usepackage[colorinlistoftodos]{todonotes}

\usepackage{times}
\usepackage{latexsym}
\usepackage{listings}

\usepackage[T1]{fontenc}

\usepackage[utf8]{inputenc}

\usepackage{microtype}

\usepackage{amsthm}

\usepackage{markdown}
\usepackage{inconsolata}

\usepackage{graphicx}

\usepackage{siunitx}
\usepackage{booktabs}

\title{MateInfoUB: A Real-World Benchmark for Testing LLMs in Competitive, Multilingual, and Multimodal Educational Tasks}

\author{Adrian Marius Dumitran, Theodor-Pierre Moroianu \and Mihnea-Vicențiu Bucă\\
        University of Bucharest \\ Faculty of Mathematics and Computer Science \\ \texttt{marius.dumitran@unibuc.ro} \texttt{\{theodor.moroianu, mihneavicentiu\}@gmail.com}}

\begin{document}

\maketitle
    
\begin{abstract}
The rapid advancement of Large Language Models (LLMs) has transformed various domains, particularly computer science (CS) education. These models exhibit remarkable capabilities in code-related tasks and problem-solving, raising questions about their potential and limitations in advanced CS contexts. This study presents a novel bilingual (English–Romanian) multimodal (text and image) dataset of multiple-choice questions derived from a high-level computer science competition. A particularity of our dataset is that the problems are conceived such that some of them are easier solved using reasoning on paper, while for others writing code is more efficient. We systematically evaluate State of The Art LLMs on this dataset, analyzing their performance on theoretical programming tasks. Our findings reveal the strengths and limitations of current LLMs, including the influence of language choice (English vs. Romanian), providing insights into their applicability in CS education and competition settings. We also address critical ethical considerations surrounding educational integrity and the fairness of assessments in the context of LLM usage. These discussions aim to inform future educational practices and policies. To support further research, our dataset will be made publicly available in both English and Romanian. Additionally, we release an educational application tailored for Romanian students, enabling them to self-assess using the dataset in an interactive and practice-oriented environment.

\end{abstract} 



\section{Introduction}
In recent years, LLMs have demonstrated revolutionary potential in natural language processing and code generation, enabling applications such as automated code writing systems and algorithmic problem-solving \cite{raihan2024largelanguagemodelscomputer, rasheed2025largelanguagemodelscode}. For instance, models like GPT-o3 exhibit remarkable proficiency in code generation and problem-solving \cite{openai2025competitiveprogramminglargereasoning}, yet their deployment in high-stakes domains remains constrained by efficiency and reliability challenges.

In the educational domain, LLMs exhibit considerable promise for enabling personalized learning and automating feedback; however, their capacity to manage complex, competition-level programming challenges—particularly in bilingual or non-English contexts—remains underexplored, with emerging critiques questioning their reliability in high-stakes scenarios, such as mathematical reasoning.  Recent analyses, such as \cite{petrov2025proofbluffevaluatingllms, mirzadeh2024gsmsymbolicunderstandinglimitationsmathematical, hendrycks2021measuringmathematicalproblemsolving} reveal that LLMs frequently produce plausible-sounding but logically flawed solutions, raising concerns about their suitability for rigorous assessments. While benchmarks like HumanEval \cite{chen2021evaluatinglargelanguagemodels, yu2024humanevalprombpppro} and MBPP \cite{austin2021programsynthesislargelanguage} evaluate general coding proficiency, they often neglect pedagogical dynamics, such as adaptive scaffolding for learners or ethical alignment with institutional values. Furthermore, studies caution that deploying LLMs in multilingual environments amplifies risks of semantic misinterpretation and cultural misalignment, necessitating rigorous scrutiny of their pedagogical robustness. \cite{rystrøm2025multilingualmulticulturalevaluating, marchisio-etal-2024-understanding}

Our work aims to address this gap by conducting a rigorous evaluation of LLMs using a bilingual dataset, thus shedding light on their strengths, weaknesses, and the nuances of language-specific performance. Our dataset is uniquely comprised of multiple-choice questions that were originally administered as part of a pre-university exam for prospective students. This setting not only simulates a high-stakes assessment environment, but also provides rich insights into the performance of LLMs on tasks that require both theoretical knowledge and practical application. Our approach allows us to identify key strengths and limitations of state-of-the-art LLMs, highlighting scenarios where additional context either bolsters performance or introduces redundancy and inefficiency. By dissecting performance variations across languages and problem types, we provide a nuanced understanding of how LLMs navigate complex educational assessments, such as those encountered in advanced computer science competitions and early university admissions. Moreover, our study raises important ethical considerations, as the use of automated assessments in educational settings must balance technological innovation with fairness and academic integrity.


Finally, to encourage further exploration and replication, the bilingual dataset\footnote{\url{https://huggingface.co/datasets/EHollower/MateInfoUB}} developed through this work will be made publicly available, offering a valuable resource for future research in both educational technology and competitive programming evaluation and an educational application\footnote{\url{https://mateinfo-ub.github.io/}} tailored for Romanian students, enabling them to self-assess using the dataset in an interactive and practice-oriented environment.

\subsection*{Main Contributions}


The main contributions of our work can be summarized as follows:
\begin{itemize}
    \item We introduce a novel \textbf{multimodal and bilingual dataset} comprising \textit{Romanian} and \textit{English}. The dataset includes \textbf{100 multiple-choice questions},  all enriched with extensive solutions in Romanian. This paper focuses specifically on benchmarking LLM performance on the Multiple Choice Question (MCQ) portion, including its multimodal aspects; the programming problems are provided as part of the dataset release for completeness and future research but are not evaluated here. We consider the evaluation of complex coding problems a distinct challenge requiring separate methodologies.


    \item Our dataset is uniquely designed so that multiple-choice problems can be solved through either mathematical and algorithmic reasoning or by generating executable Python code. Crucially, the benchmark tasks the LLMs with autonomously determining the most suitable approach—producing either direct answers or executable Python code.

    \item We provide an open-source \textbf{educational application} enabling students to interactively attempt and practice all problems included in our dataset, thereby facilitating practical engagement and learning.
\end{itemize}


\subsection*{Related Work}


The evaluation of LLMs for code generation has advanced significantly, supported by benchmarks that measure functional correctness and problem-solving capability. Seminal datasets such as HumanEval \cite{chen2021evaluatinglargelanguagemodels, yu2024humanevalprombpppro} and MBPP (Mostly Basic Python Problems) \cite{austin2021programsynthesislargelanguage} have become standard, focusing on generating standalone code from English-language prompts \cite{paul2024benchmarksmetricsevaluationscode}. While effective for assessing basic coding abilities, these benchmarks often emphasize isolated tasks, neglecting integrated reasoning, debugging, and pedagogical scaffolding \cite{fujisawa2024procbenchbenchmarkmultistepreasoning, zhang2024naturalcodebenchexaminingcodingperformance}. They also overlook ethical alignment \cite{abdulhai-etal-2024-moral}, which is critical in educational deployments.

Recent datasets attempt to address these gaps. APPS \cite{hendrycks2021measuringmathematicalproblemsolving} and CodeContests \cite{quan2025codeelobenchmarkingcompetitionlevelcode} introduce complex algorithmic problems from competitive programming, pushing models toward more advanced problem-solving. However, these datasets are monolingual and insufficiently capture linguistic diversity \cite{marchisio-etal-2024-understanding}, despite growing evidence that non-English prompts introduce semantic errors and cultural misalignment \cite{rystrøm2025multilingualmulticulturalevaluating}.

In educational contexts, systems for automated feedback \cite{Sarsa_2022} and personalized tutoring \cite{wu-hu-2023-exploring, petrov2025proofbluffevaluatingllms} rarely engage with high-stakes scenarios such as programming competitions or university admissions. This leads to concerns about fairness \cite{mouselinos2023simpleeffectiveapproachfinding}, academic integrity \cite{huang2025surveylargelanguagemodels}, and linguistic exclusion \cite{gao2024ambiguityawareincontextlearninglarge}.

Multilingual benchmarks like DS-1000 \cite{lai2022ds1000naturalreliablebenchmark} and MultiPL-E \cite{cassano2022multiplescalableextensibleapproach} broaden the scope but primarily target English programming tasks rather than bilingual educational assessments. Studies reveal that language choice affects problem comprehension \cite{moumoula2025evaluatingprogramminglanguageconfusion}, with LLMs showing systematic bias in non-English settings and often generating plausible yet logically flawed responses \cite{petrov2025proofbluffevaluatingllms, mirzadeh2024gsmsymbolicunderstandinglimitationsmathematical}. As a result, emerging frameworks call for pairing benchmarks with fairness audits \cite{du2025faircoderevaluatingsocialbias} and cultural robustness evaluations \cite{rystrøm2025multilingualmulticulturalevaluating}.

Several recent benchmarks have expanded beyond single-turn code generation to include interaction and feedback mechanisms. MINT introduces multi-turn tool use and natural language feedback \cite{wang2024mintevaluatingllmsmultiturn}, while InterCode and AppWorld emphasize coding with execution feedback and app-driven interaction \cite{yang2023intercodestandardizingbenchmarkinginteractive, trivedi2024appworldcontrollableworldapps}. SciCode curates scientific computing tasks \cite{tian2024scicoderesearchcodingbenchmark}, and XCODEEVAL targets multilingual, multitask code understanding and generation \cite{khan2023xcodeevallargescalemultilingual}. However, these benchmarks largely isolate competencies: tool use is decoupled from theoretical reasoning, and scientific or multilingual problems are rarely embedded in pedagogically structured tasks. 

Unlike existing benchmarks that focus on isolated coding tasks, our dataset integrates theoretical understanding with practical implementation through hybrid problem formats. Each item in the dataset focuses on one or more core competencies: code synthesis, mathematical reasoning, and algorithmic thinking. This flexible format mirrors the diversity of real-world computer science assessments and addresses the "theoretical blind spots" highlighted by \cite{chan2024rulebreakerschallengerevealingblind}, where language models struggle when reasoning is detached from implementation. By evaluating symbolic manipulation alongside executable code generation, our dataset offers a more comprehensive measure of educational readiness.

\section{Data Collection and Examples}


The problem set used in our study is derived from \textbf{MateInfoUB}, an annual computer science contest specifically aimed at 12th-grade students. This contest also functions as an admission exam for the Faculty of Mathematics and Computer Science at the University of Bucharest. The competition is structured into two phases:

\begin{itemize}
    \item \textbf{Phase 1:} An online round consisting of challenging multiple-choice questions. Students have access to a programming environment, but are restricted to using only publicly available resources. The use of forums, messengers, or Large Language Models (LLMs) is strictly prohibited.
    \item \textbf{Phase 2:} A live programming contest modeled after the International Olympiad in Informatics (IOI) format, featuring four programming problems. Students' solutions can earn partial points based on correctness and efficiency.
\end{itemize}

Our work exclusively focuses on the first phase of the contest, and our dataset is obtained directly from the contest organizers in Romanian, currently also available online \footnote{\url{https://mateinfo-ub.github.io/\#/toate-datele}}. Extensive solutions accompanying each problem, manually written by the authors and by undergraduate students as part of their academic practice (\textit{practică}) at the university are also available for reference and further research.

Some tasks are accompanied by an image containing a code snippet, a diagram, a graph or similar. For those tasks, we augment the statement with a clear textual description of the image's content.

Our final dataset is composed of the problems with statements and multiple choice answers in Romanian, as well as their direct translation in English. The translations are generated automatically, by using \textit{Gemini 2.0 Flash} with very strict instructions enforcing a verbatim translation. The english translations are then manually checked for correctness.

In the following, we provide two examples that illustrate the characteristics of the dataset.

\subsection*{Example: Multimodal Problem Requiring Visual Analysis}

\begin{figure}[h!]
\centering
\includegraphics[width=1\linewidth]{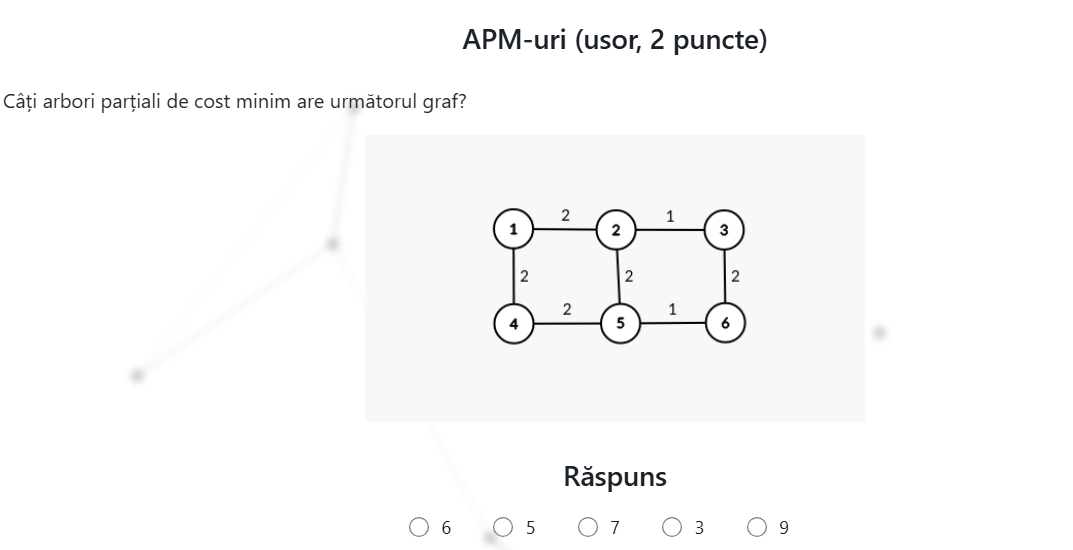}
\caption{Example multiple-choice problem requiring visual analysis (in Romanian); English translation: "AMP-uri (easy, 2 points); How many minimum spanning trees does the following graph have?"}
\label{fig:mst-example}
\end{figure}

Figure \ref{fig:mst-example} presents a typical multiple-choice question from our dataset. The problem requires determining the number of distinct Minimum Spanning Trees (MSTs) present in the provided graph. Problems of this nature are challenging for LLMs, as solutions depend significantly on visual interpretation and structural observation of the graph. Previous studies have noted similar limitations in visual reasoning tasks performed by LLMs \cite{liu2023visualinstructiontuning}.

\begin{figure}[h!]
\centering
\includegraphics[width=1.10\linewidth]{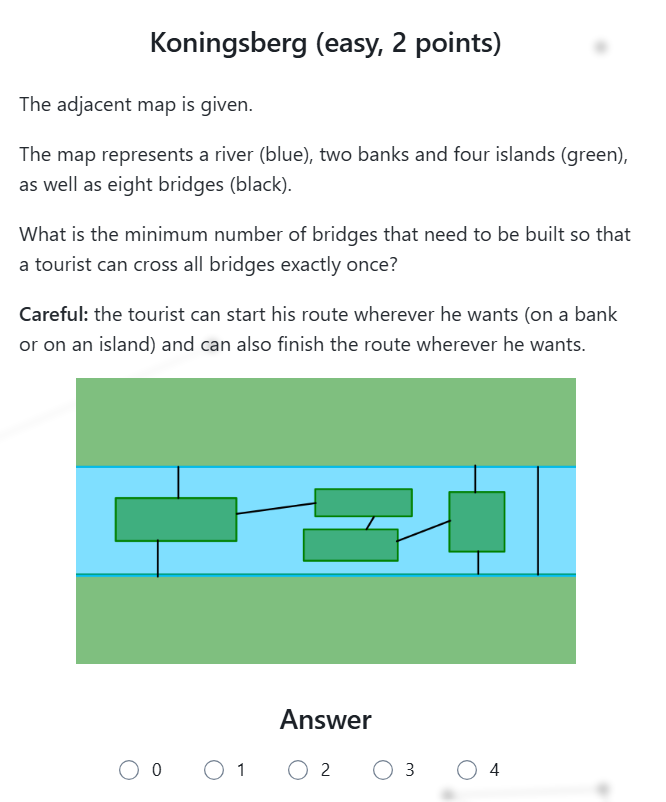}
\caption{Example multiple-choice problem requiring visual analysis (in English).}
\label{fig:koinsberg-example}
\end{figure}

Figure \ref{fig:koinsberg-example} presents another multiple choice question from our dataset. Given a map that illustrates a river with two banks and four islands linked by eight bridges, the task asks for the minimum number of additional bridges that must be built so that a tourist can cross each bridge exactly once. Problems of this nature are challenging for LLMs because they require integrating visual-spatial reasoning with graph-theoretical concepts, such as identifying Eulerian paths, which are not explicitly stated, but must be inferred from the structure of the image or diagram.

\section{Benchmarking}

In this section, we present a comprehensive overview of our benchmarking strategy designed to evaluate various aspects of Large Language Models (LLMs) performance on our bilingual, multimodal dataset. The benchmarking aims to highlight differences in performance across multiple dimensions, including language, presentation modality, availability of multiple-choice options, and problem-solving approaches.

\subsection{Methodology}


Our evaluation is based on state-of-the-art LLM models from various vendors, namely \textit{gemini-2.0-flash} and \textit{gemini-2.5-pro-exp-03-25} from \textit{Google AI Studio}, \texttt{mistral-large-latest} (April 2025) from \textit{Mistral AI}, and \textit{meta-llama/Llama-3.3-70B-Instruct-Turbo-Free}, \textit{deepseek-ai/DeepSeek-R1} and \texttt{deepseek-ai/DeepSeek-V3} from \textit{Together AI} \cite{togetherai}

We use the models via the exposed API, starting a new chat instance for each task. We provide the models with the task's statement and the multiple-choice answers. We then instruct the models to provide reasoning steps, followed by either an answer or a Python script that computes the answer.

For minimizing benchmarking frictions, we clearly provide the models with the expected output format, which resembles XML. While very forgiving, in some instances, the models fail to adhere to it (if, for instance, their answer exceeds the API's length limit). In such situations, we consider the models' answers incorrect.

\subsection{Evaluation Baseline}

We evaluate the accuracy of our models on the original tasks. Due to the multiple choice nature of the tasks, verifying the correctness of the models' solutions is trivial. We run each model on each problem 3 times, for minimizing the randomness caused by the LLMs' seed selection. We chose to use the models' default API settings (i.e., without forcing temperature to $0$) to reflect typical usage and obtain realistic levels of correctness, confidence, and creativity, as would be experienced by a standard user.

\subsection{AI vs. Human Contestants}

As our dataset comes from real contests, we compare the performance of the models with the results obtained during the \texttt{2021}, \texttt{2022}, \texttt{2023}, and \texttt{2024} editions of the contest. We evaluated the models by measuring their percentile scores compared to the results of the students who qualified for the final stage of the contest.

\subsection{Original Romanian Baseline vs. English Translations}

LLMs have been notoriously bad at reasoning in languages other than English. By comparing our baseline benchmark with the performance of LLMs on English translations of the statements, we gain insights about the model's effectiveness in a language typically underrepresented in NLP research, as opposed to English.

A comparative analysis of English and Romanian benchmarks highlights language-specific challenges and differences in LLM capabilities between languages.

\subsection{Original Multiple-Choice vs. No Multiple-Choice Variants}

We investigate how the presence or absence of multiple choice answer options affects LLM performance. By removing the multiple-choice framework, we challenge the models' capability to generate answers without guidance from predefined options.

\subsection{Chain-of-Thought vs. Direct Answer}

In our benchmarks, when prompting the models for an answer, we ask the models to provide a detailed description of the solution. The models thus respond with reasoning steps to solve the task, followed by the answer.

We measure how the performance of the models is impacted by the absence of the reasoning steps, by prompting the models to directly output the answer, without justifying it.

\subsection{Answer-only vs. Hybrid Approach}
Finally, we conduct experiments to compare LLM's performance across two different reasoning strategies:
\begin{itemize}
    \item \textbf{Hybrid approach}: The model autonomously chooses whether to solve the problem via code generation or direct reasoning (our baseline).
    \item \textbf{Think-only}: The model is restricted to providing direct theoretical or conceptual solutions, without the possibility of running \textit{python} code.
\end{itemize}

We do not consider the third option (forcing the model to produce \textit{python} code), as we experimentally see the model can write a trivial script printing a hard-coded answer, making the experiment uninteresting.

\section{Results}

In this section, we present the findings of our benchmarking evaluations.

Overall, our analyses suggest significant variations in LLMs performance across different scenarios.

We acknowledge that the outreach of benchmarks might be limited by the size relatively small of our dataset, but our measurements suggest the following trends:  

\begin{itemize}
    \item \textbf{Language Comparison:} Our measurements indicate various differences in model accuracy and problem-solving capabilities when problems are presented in Romanian versus English, with some models benefiting from a verbatim translation of the statements to English, a language they are more familiar with, while and others perform better when exposed to the original statements. 
    \item \textbf{Multiple-choice Contexts:} We observe that the availability of multiple-choice options slightly improves model accuracy compared to scenarios where these options are not provided.
    \item \textbf{Reasoning Strategies:} Benchmarks indicate a promising performance of hybrid strategies, where models autonomously select between code generation and direct reasoning, outperforming exclusive reasoning approaches.
    \item \textbf{Human Performance Comparison}: Comparing the performance of models with the performance of high school students taking part in the contest, we find that newer models consistently outperform most students.
    \item \textbf{Breadth of Capabilities:} Models demonstrate a broader range of problem-solving skills than originally anticipated, effectively addressing a very diverse set of exercises. In particular, of the $100$ exercises in our dataset, all but three were solved by at least one model, highlighting their versatility and adaptability to different types of problems.
\end{itemize}

Detailed findings from each of our experiments are discussed in the following sections.

\subsection{Baseline and comparison with human rankings}

We find models quite capable, with newer models capable of solving most tasks. On average, models achieve a difficulty-weighted accuracy of \textit{52\%} (easy problems are worth 2 points, medium problems 3 points, and hard problems 5 points). A complete breakdown of the accuracy of the models is available in Table \ref{tab:performance_metrics_ro}.

\begin{table}[h!]
\centering
\small
\resizebox{\columnwidth}{!}{
\begin{tabular}{lSSS}
\toprule
\textbf{Model} & \textbf{Easy (\%)} & \textbf{Medium (\%)} & \textbf{Hard (\%)} \\
\midrule
Gemini 2.5 Exp & \textbf{96.7} & \textbf{85.6} & \textbf{76.7} \\
Gemini 2.0 Flash & 71.3 & 60.0 & 35.0 \\
Llama 3.3 70B  & 51.3 & 31.1 & 18.3 \\
DeepSeek R1 & 54.0 & 34.4 & 21.7 \\
DeepSeek V3 & 72.0 & 63.3 & 30.0 \\
Mistral Large  & 62.0 & 43.3 & 31.7 \\
\bottomrule
\end{tabular}
}
\caption{Performance metrics for different models across difficulty levels in Romanian (larger is better).}
\label{tab:performance_metrics_ro}
\end{table}

One can see that \textit{Gemini 2.5}, a reasoning-focused model, outperforms all others, including \textit{DeepSeek R1}. However, upon closer examination, we found that \textit{DeepSeek R1} frequently produces answers that exceed the API's maximum response length, leading to truncation and, consequently, incorrect outputs.

In some cases, such as when attempting to manually solve complex counting problems, the model's output becomes excessively long. We consider it most appropriate to adhere to the API vendor’s configured maximum response length and treat truncated or incomplete answers as incorrect.

As we have data on the scores obtained by students qualified in the \textit{2021, 2022, 2023} and \textit{2024} editions of the contest, we can compute the percentile (i.e. the percentage of students doing better than the model) of the qualified students. The results are available in Table \ref{tab:percentiles_years}.

\begin{table}[h!]
\centering
\small
\resizebox{\columnwidth}{!}{
\begin{tabular}{lSSSS}
\toprule
\textbf{Model} & \textbf{2021} & \textbf{2022} & \textbf{2023} & \textbf{2024} \\
\midrule
DeepSeek V3 & 38.22 & 27.12 & 26.15 & 9.95 \\
Gemini 2.5 Exp & 0.64 & 1.69 & 0.51 & 0.00 \\
Gemini 2.0 Flash & 50.96 & 55.93 & 56.41 & 1.05 \\
Llama 3.3 70B & 100.00 & 100.00 & 100.00 & 100.00 \\
DeepSeek R1 & 100.00 & 100.00 & 79.49 & 100.00 \\
Mistral Large & 100.00 & 92.09 & 85.64 & 9.95 \\
\bottomrule
\end{tabular}
}
\caption{Average percentiles of models compared to real students across different years (smaller is better).}
\label{tab:percentiles_years}
\end{table}

Models show a constant improvement over the years, which we theorize can be explained by a combination of the following hypotheses:
\begin{itemize}
    \item Older contests have more ad-hoc problems, which models tend to struggle with.
    \item Starting in 2022, the UK stopped its financial support for EU students, including Romania. Thus, many students exploring alternative opportunities took part in the contest. As alternative abroad universities grew in popularity among high school students, interest in the contest could have decreased.
    \item The student participating in the contest in 2022, 2023 and 2024 were the most impacted by the \textit{Covid-19} remote studying mandates during early high school.
\end{itemize}

We strongly believe that our dataset is not tainted (i.e., that models were not trained on it). While the statements were publicly available before we started our research, we are the first to compile a dataset that maps these problems to their corresponding answers.

In other words, while it is plausible that model training corpora may have included the raw statements, the solutions could not have been included, since all tasks are original, and our dataset is the first to pair them with verified multiple-choice answers.

\subsection{Original language vs. English translation}


Our experiments show that most models perform better on the Romanian version of the questions than on the English one. This gap likely arises from two factors. First, our English statements are verbatim translations of the original Romanian text, which can lose nuance and clarity and introduce artifacts (translationese) that impair model understanding.

Second, since the raw Romanian problems were publicly available before our work, it is plausible that models encountered those during training, whereas our English translations are novel; thus, they effectively function as a partial unseen validation set. We therefore consider the benchmark bilingual, while acknowledging that translation quality and prior exposure may both contribute to the observed performance drop.

The \textit{DeepSeek} family of models sees a $10\%$ gain in accuracy when solving the English variant, suggesting reduced multilingual abilities of the models.

The results of the experiment are available in Table \ref{fig:multiple-choice-vs-not}.

\begin{table}[h!]
\centering
\small
\begin{tabular}{lSS}
\toprule
\textbf{Model} & \textbf{English (\%)} & \textbf{Romanian (\%)} \\
\midrule
DeepSeek V3 & 0.61 & 0.55 \\
Gemini 2.5 Exp & 0.84 & 0.86 \\
Gemini 2.0 Flash & 0.50 & 0.55 \\
Llama 3.3 70B & 0.33 & 0.34 \\
DeepSeek R1 & 0.52 & 0.37 \\
Mistral Large & 0.43 & 0.46 \\
\midrule
\textbf{Overall Average} & 0.54 & 0.52 \\
\bottomrule
\end{tabular}
\caption{Average scores for models across different languages.}
\label{tab:average_scores}
\end{table}

\subsection{Multiple-choice options provided vs. not provided}

\begin{figure*}[!t]
\centering
\includegraphics[width=1\linewidth]{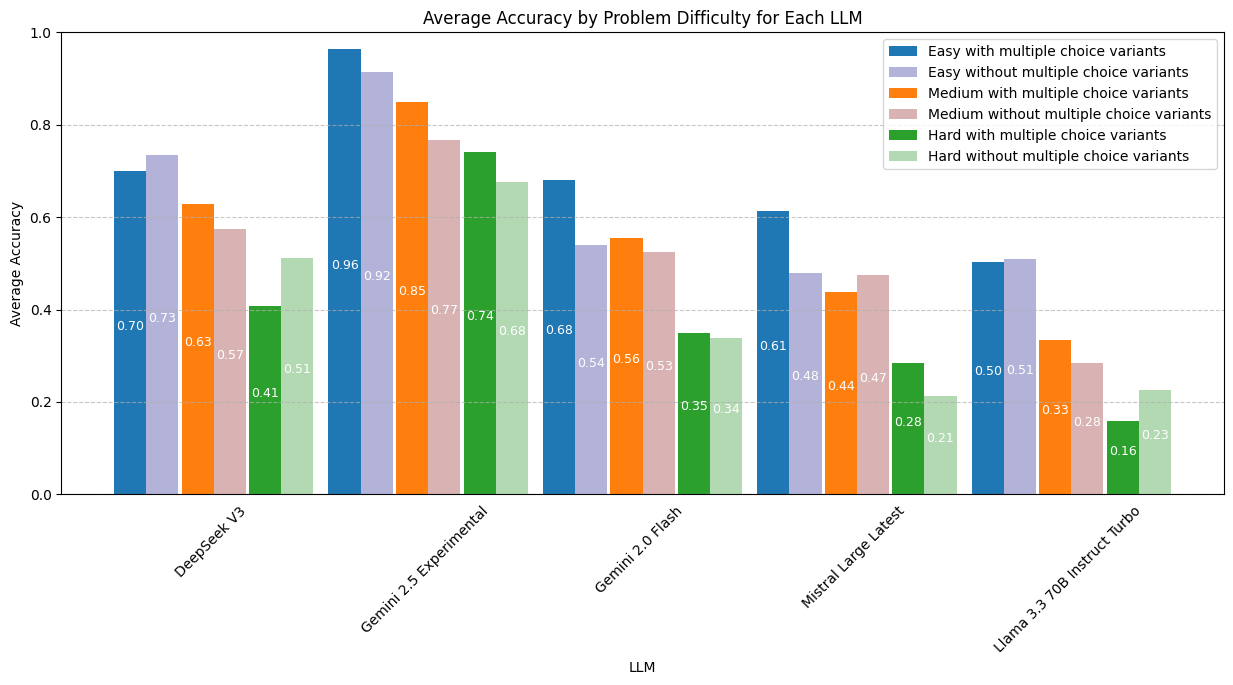}
\caption{Performance of models when the multiple choice options are provided vs when they are not.}
\label{fig:multiple-choice-vs-not}
\end{figure*}

We observe a slight decline in performance when the multiple choice variants are not provided, which aligns with the contests' design goals of making the variants unhelpful for solving the tasks. Considering that models tend to \textit{guess} an answer and hallucinate a justification when they cannot solve the task, we believe that the difference is caused by models having a higher chance of \textit{guessing} the correct answer. The results can be seen in Figure \ref{fig:multiple-choice-vs-not}.

\subsection{Chain-of-Thought vs. Direct Answer}

We observe a slight decline in performance when models are only prompted for the answer, as opposed to first providing a justification, or reasoning. While we expect \textit{reasoning} models like \textit{Gemini 2.5 Exp} and \textit{DeepSeek-R1} to be invariant to the change (due to their own reasoning process), \textit{DeepSeek-R1}'s internal chain-of-thought reasoning increases in length, which causes some of its answers to be truncated and invalidated. The full results are available in Table \ref{tab:accuracy_comparison_reasoning}.

\begin{table}[h!]
\centering
\small
\begin{tabular}{lSSSS}
\toprule
\textbf{Model} & \textbf{Easy} & \textbf{Medium} & \textbf{Hard} & \textbf{Average} \\
\midrule
\multicolumn{5}{c}{\textbf{With Reasoning}} \\
\midrule
DeepSeek V3 & 0.70 & 0.63 & 0.41 & 0.58 \\
Gemini 2.5 Exp & 0.96 & 0.85 & 0.74 & 0.85 \\
Gemini 2.0 Flash & 0.68 & 0.56 & 0.35 & 0.53 \\
Llama 3.3 70B & 0.50 & 0.33 & 0.16 & 0.33 \\
DeepSeek R1 & 0.65 & 0.43 & 0.26 & 0.45 \\
Mistral Large & 0.61 & 0.44 & 0.28 & 0.45 \\
\midrule
\textbf{Overall Average} & & & & 0.53 \\
\midrule
\multicolumn{5}{c}{\textbf{Without Reasoning}} \\
\midrule
DeepSeek V3 & 0.73 & 0.54 & 0.48 & 0.58 \\
Gemini 2.5 Exp & 0.95 & 0.81 & 0.79 & 0.85 \\
Gemini 2.0 Flash & 0.56 & 0.50 & 0.11 & 0.39 \\
Llama 3.3 70B & 0.53 & 0.27 & 0.19 & 0.33 \\
DeepSeek R1 & 0.28 & 0.12 & 0.00 & 0.13 \\
Mistral Large & 0.48 & 0.35 & 0.15 & 0.33 \\
\midrule
\textbf{Overall Average} & & & & 0.43 \\
\bottomrule
\end{tabular}
\caption{Comparison of average scores with and without reasoning for various models.}
\label{tab:accuracy_comparison_reasoning}
\end{table}

\subsection{Answer-only vs. Hybrid Approach}



\begin{figure*}[h!]
\centering
\includegraphics[width=\linewidth]{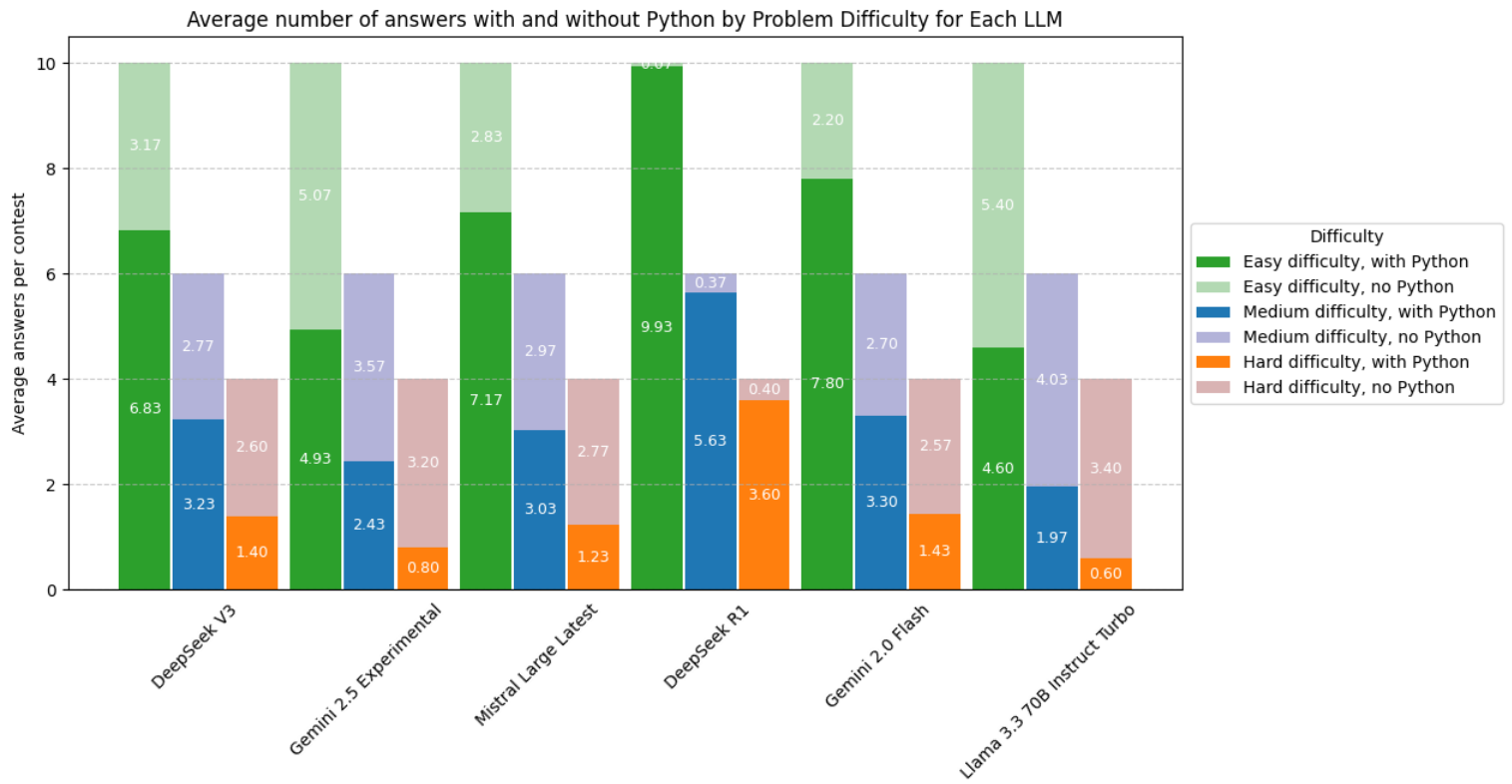}
\caption{Percentage of direct answers vs \textit{python} answers when the models are given the choice between the two. Columns are scaled the number of tasks of each difficulty (10 easy, 6 medium, 4 hard).}
\label{fig:plot_python_vs_answers_optional}
\end{figure*}

In Figure \ref{fig:plot_python_vs_answers_optional}, we can see the distribution of \textit{python} answers and direct answers, when the models can choose unconstrained how to provide an answer for a given task (i.e., models can freely pick between providing a direct answer and providing a \textit{python} code).

When \textit{Python} is no longer allowed, the models perform unexpectedly worse, as more than half of their answers rely on executing \textit{Python} code. We run the experiment on a subset of the models, and the results are available in Table \ref{tab:comparison_scores_no_python}. 

\begin{table}[h!]
\centering
\small
\begin{tabular}{lSSS}
\toprule
\textbf{Model} & \textbf{Easy} & \textbf{Medium} & \textbf{Hard} \\
\midrule
\multicolumn{4}{c}{\textbf{Python Code Allowed}} \\
\midrule
DeepSeek V3 & 0.70 & 0.63 & 0.41 \\
Gemini 2.0 Flash & 0.68 & 0.56 & 0.35 \\
\midrule
\multicolumn{4}{c}{\textbf{Python Code Not Allowed}} \\
\midrule
DeepSeek V3 & 0.65 & 0.48 & 0.28 \\
Gemini 2.0 Flash & 0.59 & 0.43 & 0.20 \\
\bottomrule
\end{tabular}
\caption{Comparison of average scores for DeepSeek V3 and Gemini 2.0 Flash with and without Python code.}
\label{tab:comparison_scores_no_python}
\end{table}

\subsection{Discussion of Benchmark Results}

Our findings have two key implications for computer‑science education. First, assessments should combine tool‑enabled tasks with those requiring scaffolding and manual reasoning to accurately gauge student mastery. Second, instructors and contest organizers should monitor for anomalous solution patterns—such as perfectly formatted code or implausibly high confidence scores—to detect unauthorized LLM use.  

We also provide descriptive plots covering the experiments contained in our research. They are available in Appendix \ref{sec:experiment-plots}.

\section{Application}

In parallel with our experiments, and inspired by our dataset, we develop a web-based application, which can be used as a training ground of students looking to compete in the contest.

The application is freely accessible online  \footnote{\url{https://mateinfo-ub.github.io/}} and, among others, allows students to:

\begin{itemize}
    \item Preview the statements of all editions of the contest.
    \item Simulate an edition of the contest.
    \item Automatically grade their attempt.
\end{itemize}

The application is implemented in \textit{React}, and is hosted on \textit{Github Pages}. Due to its limited functionalities, it does not require any kind of dynamic backend, and all of its assets, including statements, solutions, and images, can be packaged statically, making deployment easier.

Screenshots of the application and a description of its functionalities are available in the Appendix \ref{sec:appendix-mateinfo-ub-website}.

\section{Future Work}

Several avenues for further research are highlighted by our preliminary findings and current limitations. Key directions include:

\begin{itemize}
    \item \textbf{Expanded Benchmarking:} Conducting extensive experiments involving additional competitive programming datasets (potentially using cross-validation) and further expanding multilingual analyses. This could also involve analyzing model performance over varying levels of intrinsic \textbf{problem complexity}.

    \item \textbf{Live Contest-Based Evaluation:} Introducing the second phase featuring a plethora of programming contests modeled after the International Olympiad in Informatics (IOI), with multiple problems graded on partial correctness and efficiency. This would enable a deeper analysis of LLMs' algorithmic reasoning and problem-solving capabilities in a structured, task-oriented environment.

    \item \textbf{Fine-Tuning and Contextual Support:} Investigating the impact of fine-tuning LLMs on domain-specific data, leveraging RAG methods, or exploring the effect of providing incremental \textbf{contextual hints or scaffolding} to guide model reasoning.

    \item \textbf{Model Efficiency and Scalability:} Exploring methods to optimize model inference times and computational efficiency for real-world educational deployment.

    \item \textbf{Enhanced Ethical Solutions:} Developing and evaluating robust technological and educational solutions that address challenges of academic integrity related to the use of LLM.
\end{itemize}

Pursuing these directions can deepen the understanding of LLM capabilities and limitations, contributing to their sustainable and ethical integration in education.

\section*{Limitations}

Although our study provides valuable insights into Large Language Models' (LLMs) performance on bilingual educational assessments, several limitations must be acknowledged.

First, although our dataset features a diverse set of problems from a high-stakes computer science competition, the scope remains limited to the Romanian educational context. Generalization of our findings to other linguistic or educational settings may require additional validation.



Second, our dataset and benchmarks currently focus primarily on the immediate accuracy of LLM-generated solutions. Future work should explore complementary evaluation metrics, including efficiency, robustness to variations in problem presentations, and detailed error analyses, which would provide deeper insights into model performance and reliability in educational contexts.

Lastly, our benchmarking has not explored the impact of methods such as retrieval-augmented generation (RAG) or fine-tuning of LLMs. Future work incorporating these approaches could reveal further improvements in performance and greater adaptability to specific educational tasks and datasets.

\section*{Ethical Considerations}

A central motivation for this study is assessing the current capabilities of Large Language Models (LLMs) due to significant ethical challenges posed by their increasing accessibility during online assessments, particularly in competitive contexts such as MateInfoUB. Our benchmarking explicitly aims to identify tasks and problem structures that LLMs struggle to solve reliably. The insights gained allow educators and contest organizers to structure future contests in a way that mitigates unfair advantages gained through unauthorized LLM use.

Although our current findings suggest it remains possible to maintain fairness in online competitions for now by emphasizing problems that LLMs find challenging, this strategy will likely become less effective as LLM capabilities rapidly improve. Therefore, it is increasingly important for educational institutions and competition organizers to proactively adopt technical solutions designed to uphold academic integrity. Such solutions could include software capable of capturing contestant screens, monitoring interactions, and verifying participant authenticity. Additionally, educational efforts should emphasize ethical awareness and responsible technology use, preparing students to navigate the evolving landscape of educational assessments responsibly.

At the same time, we acknowledge that releasing a dataset modeled after real pre-university exams introduces the risk of misuse, particularly fine-tuning LLMs to artificially boost exam performance without genuine understanding. Our benchmark is intended for controlled research and diagnostic evaluation, not as training material for high-stakes testing. Responsible use requires avoiding practices that could compromise the integrity and fairness of educational assessments.



\bibliography{custom}

\appendix
\onecolumn

\section{Online Training Platform}
\label{sec:appendix-mateinfo-ub-website}

\begin{figure}[h!]
\centering
\includegraphics[width=0.9\linewidth]{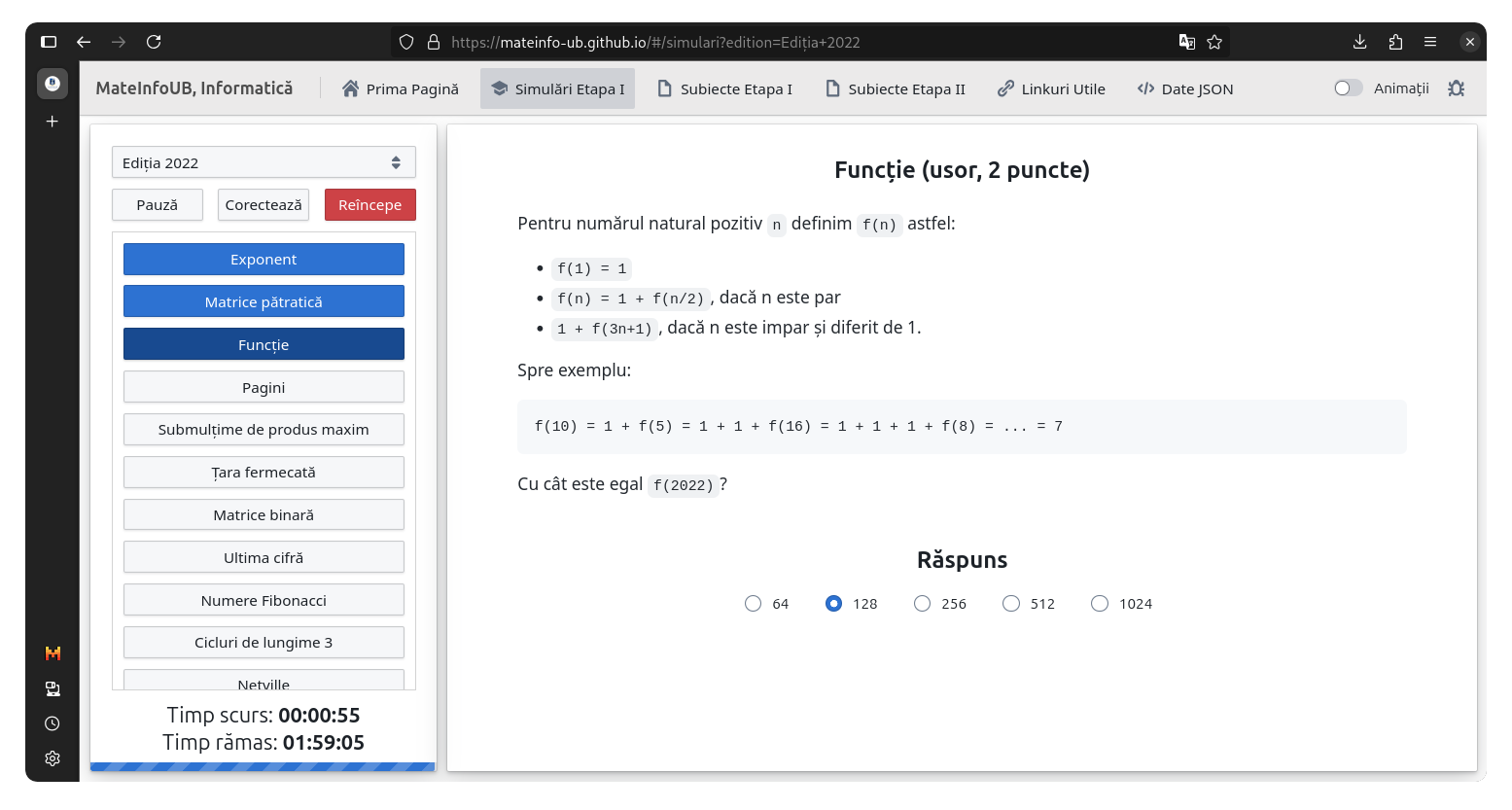}
\caption{Screenshot of the web application while solving a contest.}
\label{fig:website-mateinfo-ub-solving}
\end{figure}

\begin{figure}[h!]
\centering
\includegraphics[width=0.9\linewidth]{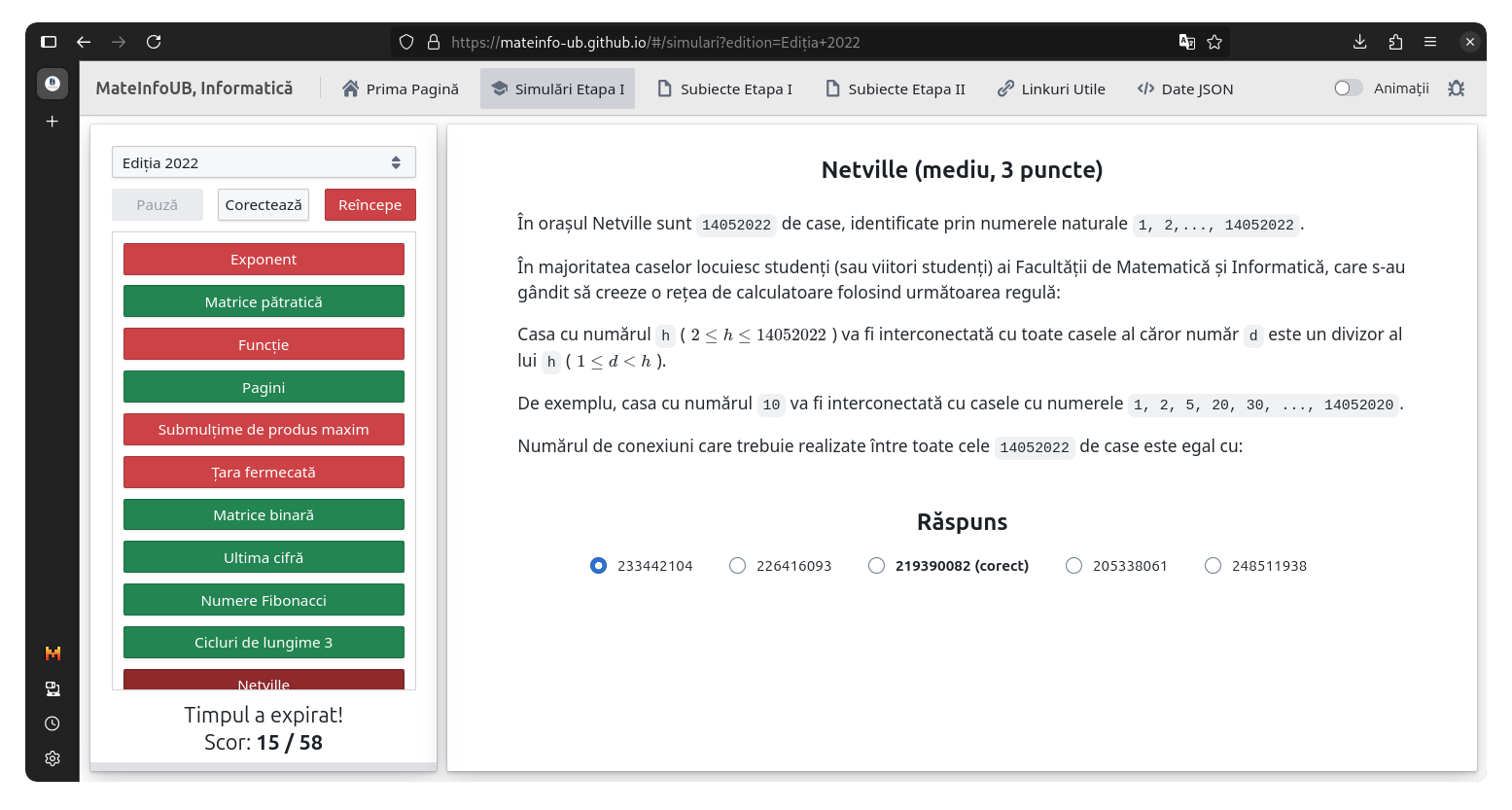}
\caption{Screenshot of the web application correcting an attempt.}
\label{fig:website-mateinfo-ub-grading}
\end{figure}

In the image \ref{fig:website-mateinfo-ub-solving} one can see the user interface of the application during a simulation of the 2022 edition of the contest, and the image \ref{fig:website-mateinfo-ub-grading} shows the user interface after the contest's timer ends or the user manually stops it.

While simple, the application contains all the necessary features for an exam-like environment:

\begin{itemize}
    \item A timer.
    \item A menu with all of the problems of the contest, ordered by difficulty and color-coded based on the answer provided (blue during the contest and green / red afterwards).
    \item A problem viewer, where users can read the statements and provide answers.
\end{itemize}

In addition to the simulation page, the application contains pages with the \textit{pdf} statements, exactly as they were during the corresponding exams, and links to useful resources.

\section{Experiment Plots}
\label{sec:experiment-plots}

In this section, we present visualizations of our experimental results.

\begin{figure*}[h!]
\centering
\includegraphics[width=\linewidth]{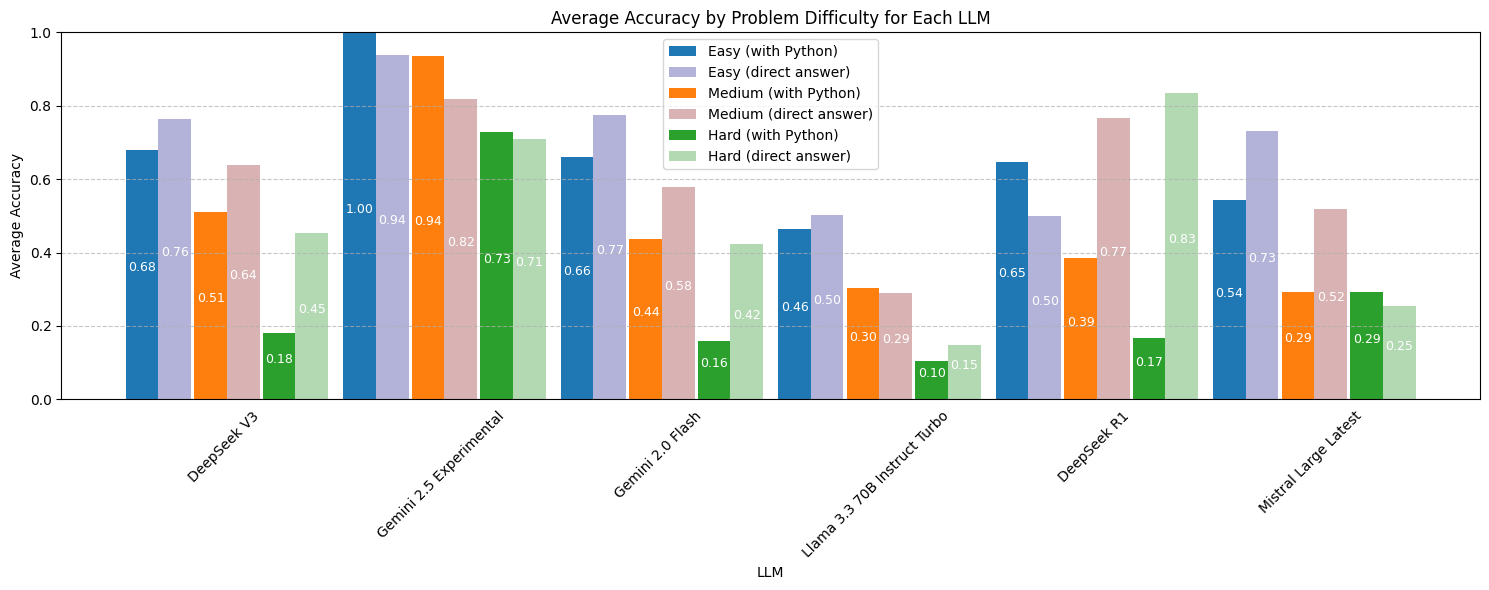}
\caption{Accuracy of the unconstrained \textit{Python} and direct answers of the models.}
\label{fig:unconstrained_python_no_python}
\end{figure*}

In Figure \ref{fig:unconstrained_python_no_python}, we plot the accuracy of the models when providing an answer as \textit{Python} code or as a direct answer. As the models can freely choose how to answer, we can see some interesting trends. For example, on hard problems, models are significantly more likely to get the right answer when providing \textit{Python} code.

\begin{figure*}[h!]
\centering
\includegraphics[width=\linewidth]{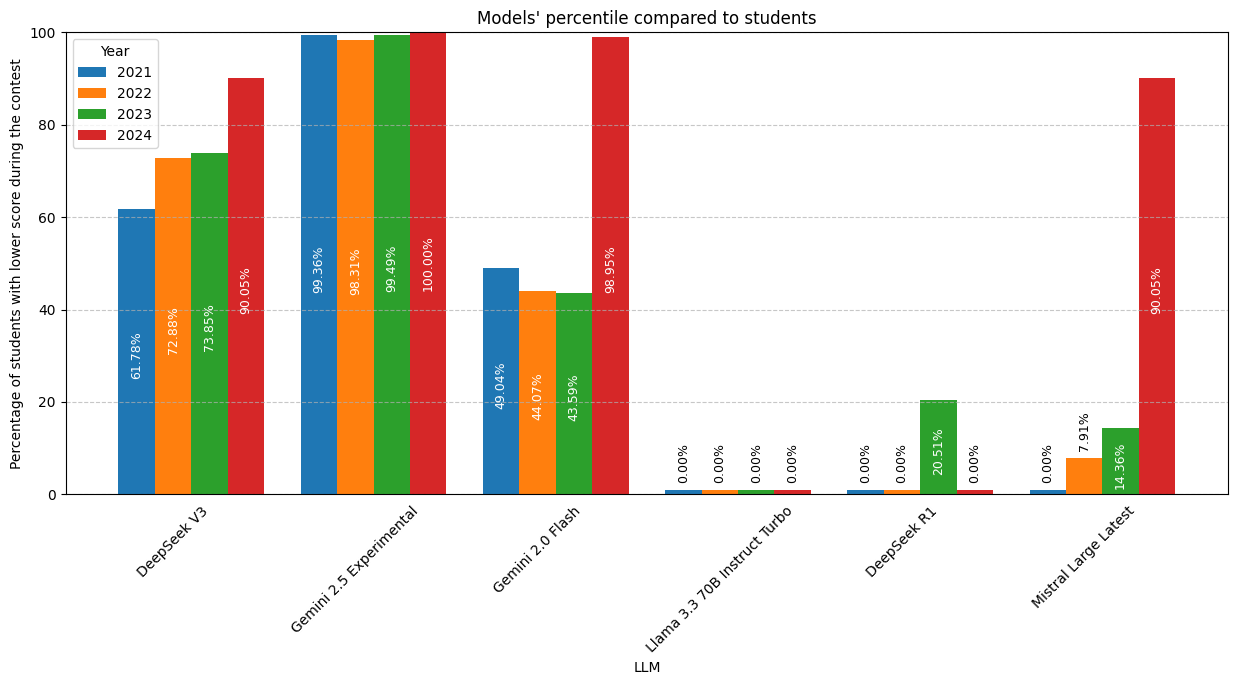}
\caption{Percentage of students with a lower score than the models, by year.}
\label{fig:models_vs_studends_plot}
\end{figure*}

In Figure \ref{fig:models_vs_studends_plot}, we plot the percentile of the models when comparing their score with the scores of students advancing to the next phase of the contest, by year. For instance, \textit{Gemini 2.5 Exp} ranks first for 3 out of the 4 years, while \textit{Llama 3.3} ranks last all years.

\begin{figure*}[h!]
\centering
\includegraphics[width=\linewidth]{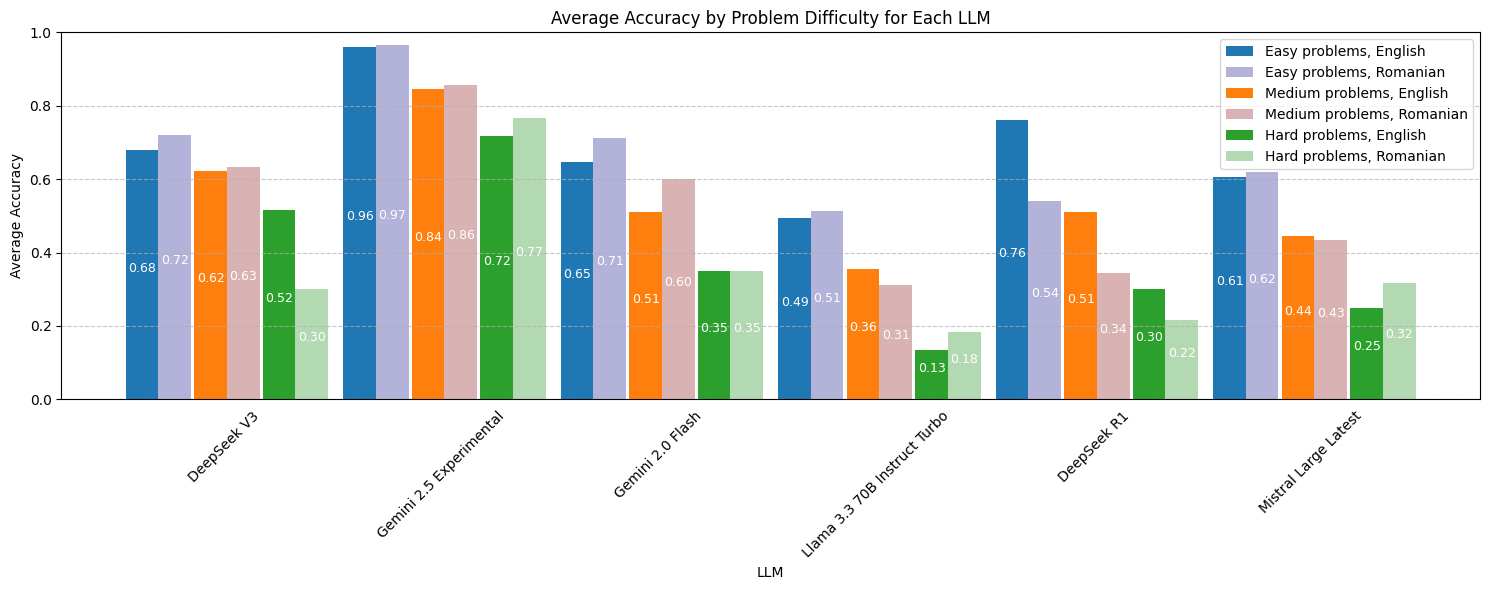}
\caption{Performance of the models by language}
\label{fig:english_vs_romanian}
\end{figure*}

Figure \ref{fig:english_vs_romanian} plots the accuracy of models by language. Except for \textit{DeepSeek-R1}, models tend to achieve a higher score in Romanian, the original language of the tasks.

When prevented to 

\begin{figure*}[h!]
\centering
\includegraphics[width=\linewidth]{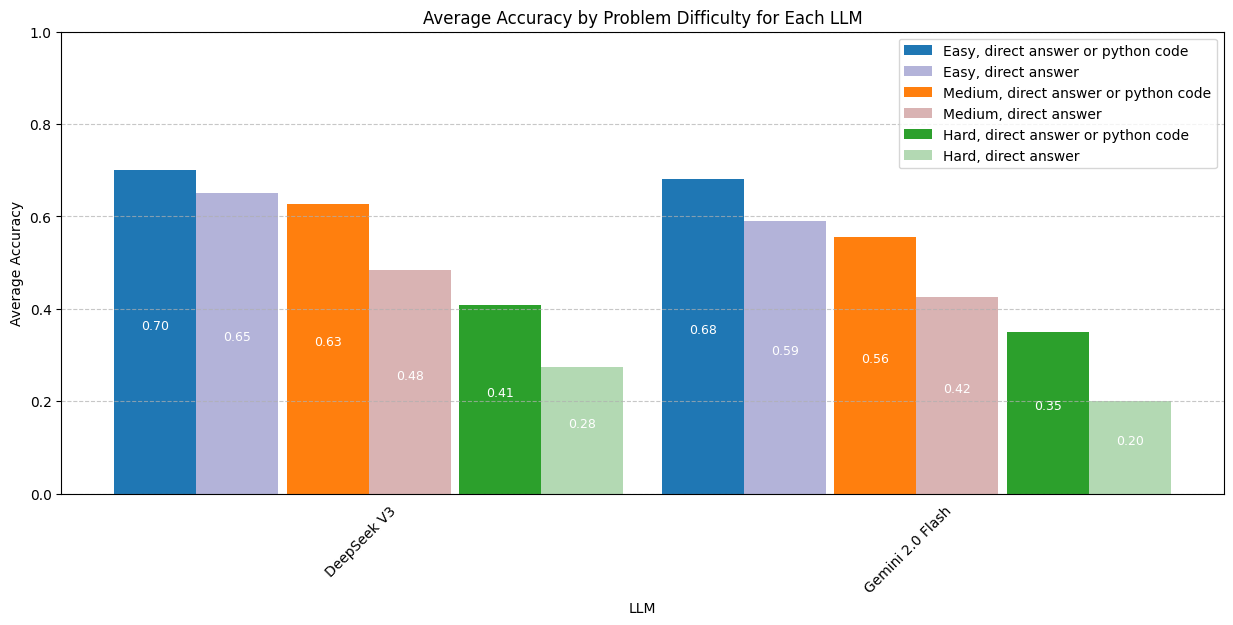}
\caption{Performance of the models when they are allowed to produce \textit{Python} code and when they have to provide the answer directly.}
\label{fig:python_allowed_vs_not_allowed}
\end{figure*}

\end{document}